\documentclass[aps,showpacs,preprintnumbers,amsmath,amssymb,amsfonts,superscriptaddress,floatfix]{revtex4}

\usepackage[T1]{fontenc}
\usepackage[latin1]{inputenc}
\usepackage{graphics}
\usepackage{color}

\usepackage{amssymb}
\usepackage{amsmath}
\usepackage{overpic}
\usepackage{epstopdf}
\usepackage{bm}
\usepackage{graphicx}
\usepackage{braket}

\newcommand{\be}{\begin{equation}}
\newcommand{\ee}{\end{equation}}
\newcommand{\bea}{\begin{eqnarray}}
\newcommand{\eea}{\end{eqnarray}}

\def\th{\theta}

\def\lb{\label}

\newcount\bozza \bozza=1
\ifnum\bozza=1
\newdimen\shift \shift=-2truecm
\def\lb#1{%
{\label{#1}\rlap{\kern\shift{$\scriptstyle#1$}}}}
\else\def\lb#1{\label{#1}} \fi

\begin{document}

\title{BKT universality class in the presence of correlated disorder}

\author{I. Maccari}
\affiliation{ISC-CNR and Dept. of Physics, Sapienza University of Rome,
  P.le A. Moro 5, 00185, Rome, Italy}
\author{C.Castellani}
\affiliation{ISC-CNR and Dept. of Physics, Sapienza University of Rome,
  P.le A. Moro 5, 00185, Rome, Italy}
\author{L.~Benfatto} 
\affiliation{ISC-CNR and Dept. of Physics, Sapienza University of Rome,
  P.le A. Moro 5, 00185, Rome, Italy}

\date{\today}

\begin{abstract}

The correct detection of the Berezinskii-Kosterlitz-Thouless (BKT) transition in quasi-two-dimensional superconductors still remains a controversial issue. Its main signatures, indeed, are often at odd with the theoretical expectations. In a recent work\cite{me} we have shown that the presence of spatially correlated disorder plays a key role in this sense being it the reason underlying the experimentally-observed smearing of the universal superfluid-density jump. In the present paper we closely investigate the effects of correlated disorder on the BKT transition, addressing specifically the issue whether it changes or not the BKT universality class.

\end{abstract}
\maketitle


\section{Introduction}

\begin{figure}[h!]
\centering
\includegraphics[width=0.7\linewidth]{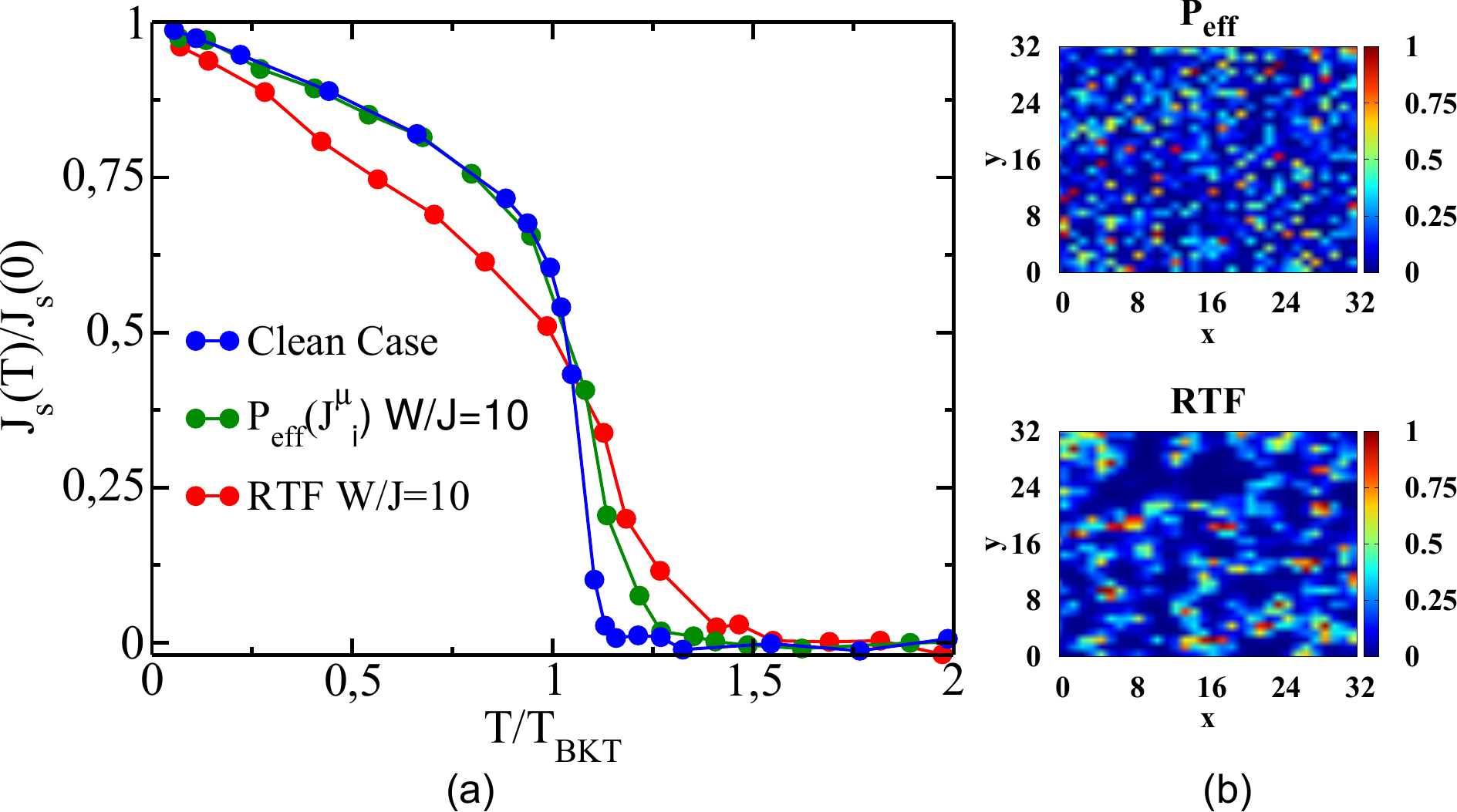}
\caption{(a) Rescaled curves of the superfluid stiffness $J_s(T)$ by its zero-temperature value $J_s(T=0)$ for the clean case, the uncorrelated $P_{eff}$ and correlated RTF disordered case at $W/J=10$. The temperature axis has been rescaled too by the value of the BKT critical temperature obtained from the intersection between the critical line $2T/\pi$ and the superfluid-stiffness itself\cite{me}. Despite the strong disorder the $P_{eff}$ curve shows only a small finite-size effect above $T_c$, while the RTF stiffness is dramatically modified above and below the transition.(b)Maps of the couplings $J_{i,i+x}$ for both the spatially uncorrelated ($P_{eff}$) and correlated (RTF) disorder. The disorder level is taken here fixed to $W/J=10$ while the linear size of the system is $L=128$.}
\label{intro}
\end{figure}

\noindent
Despite its age, the Berezinskii\cite{berezinsky}-Kosterlitz and Thouless\cite{KT, K} (BKT) transition still constitutes a very active field of research from both an experimental and a theoretical perspective.
Actually, a wide  range of phenomena belongs to its universality class: from the quantum metal-insulator transition in one dimension to the Coulomb-gas screening transition in 2D, and of course the metal-to-superfluid transition in 2D\cite{review}.
Since its first experimental detection in He films\cite{He1984}, the BKT transition has been investigated in many different real systems such as cold-atoms systems made of bosons\cite{dalibart_nature06}, neutral fermions\cite{murthy_prl15} and also in quasi-two-dimensional (2D) superconductors (SC).
The latter case includes not only thin films of conventional \cite{fiory_prb83}-\cite{Mondal_2011} and unconventional\cite{Bonn}-\cite{popovic_prb16} superconductors, but also artificially confined 2D electron gas at the interface between two insulators in artificial heterostructures\cite{bert_prb12,bid_prb16}, or in the top-most layer of ion-gated superconducting (SC) systems\cite{iwasa_science15}.
However, the experimental observations made so far on 2D superconductors have raised new questions on the nature of the transition occurring in such systems being them often at odd with the BKT theoretical predictions.
A typical example is the temperature dependence of the superfluid density which gives access to the most spectacular signature of the BKT transition: as soon as the system reaches the BKT critical temperature ($T_{BKT}$), the proliferation of free topological defects (vortices) within the system leads the superfluid density to jump suddenly to zero causing, at the same time, the vanishing of the SC state. Nonetheless, as reported by several experimental results\cite{fiory_prb83,lemberger_prl00,armitage_prb07,armitage_prb11,mondal_bkt_prl11,goldman_prl12,yazdani_prl13,Kamlapure_2010}, such sharp jump at the critical temperature results to be significantly smeared out around $T_{BKT}$, revealing a smooth downturn definitely broader than what observed in the case of superfluid helium films\cite{He1984}. This effect is even more dramatic in ultrathin films of cuprate superconductors\cite{lemberger_prb12}, where the BKT jump is completely lost by underdoping. 
Beyond the well known differences between superfluid and superconducting systems, the latter ones exhibit as common characteristic a pronounced spatial inhomogeneity of the SC order parameter. 
Such SC-state fragmentation, which occurs on a mesoscopic scale, can be due to the presence of strong disorder, as for thin disordered films of conventional superconductors, to the artificial optical confinement, as in the SC interfaces, or to the intrinsic nature of the system, as it occurs in cuprate superconductors.
Indeed, as shown theoretically\cite{trivedi_prb01,dubi_nat07,ioffe,nandini_natphys11,seibold_prl12,lemarie_prb13}, the formation of a ``granular'' inhomogeneous SC state is the way out of superconductivity, which requires phase coherence, to survive in the presence of disorder-induced charge localisation.
From these observations, it comes natural to wonder whether the observed broadening of the BKT transition can be due to the presence of such spatially correlated disorder within the system.\\
In a recent paper\cite{me}, we have addressed this interesting issue by means of Monte Carlo simulations on the classical 2D XY model:

\be
H_{XY}=-\sum_{\braket{i,j}} J_{ij} \cos(\th_i -\th_j),
\label{Hxy}
\ee 
where $\theta_i$ models the SC phase and $J_{ij}$ are the random couplings between neighbouring sites $i,j$, mimicking the random Josephson-like couplings between coarse-grained adjacent SC islands.
The granular inhomogeneity of the SC order parameter is thus embedded in the couplings $J_{ij}$, whose disordered structure has been generated by the mean-field solution of the (quantum) $XY$ pseudo-spin 1/2 model in random transverse field (RTF)\cite{lemarie_prb13,Cea,ma_prb85, me}:
\begin{equation}
\label{hamps}
\mathcal{H}_{PS}\equiv -2\sum_i\xi_iS_i^z-2J\sum_{\langle i,j\rangle}\left(S^+_iS^-_j+h.c.\right),
\end{equation}
recently proven to model disordered superconductors with a non-trivial space structure\cite{ioffe, lemarie_prb13,Cea}.
In the pseudo-spin language $S^{z} = \pm 1/2$ corresponds to a site occupied
or unoccupied by a Cooper pair, while superconductivity corresponds to
a spontaneous in-plane magnetization, e.g. $\langle S^x_i\rangle \neq 0$, controlled by the coupling $J$. The random transverse field $\xi_i$, box distributed between $-W$ and $W$, simulates the effect of disorder, which tends to localize the Cooper pair on each site or, in terms of spins, to align them out of the $x$-$y$ plane. This competition is well captured by the mean-field solution of the model \eqref{hamps}, obtained by determining the value of the in-plane local magnetization $\langle S^x_i\rangle$. While at small $W/J$ $\langle S^x_i\rangle\simeq 1/2$ everywhere, as $W/J$ increases the pseudo-spins partly orient along the $\hat{z}$ direction suppressing the in-plane component, i.e. the local value of the SC order parameter. It can also be shown\cite{Cea,me} that the SC phase fluctuations on top of this granular SC ground state are controlled by an inhomogeneous local stiffness  $J_{ij}=J\langle S^x_i\rangle\langle S^x_j\rangle$  between  neighboring $i,j$ sites.\\ 
In the following we will make use exactly of such inhomogeneous local stiffness $J_{ij}$ as couplings constants for the classical  disordered 2D XY model\eqref{Hxy}. In particular, we will refer to these spatially-correlated disordered couplings as RTF, while with $P_{eff}$ we will indicate spatially uncorrelated couplings extracted randomly from the same probability distribution which represents the RTF maps.
Finally, the label $W/J$ will indicate the disorder level considered(see \cite{me} for more details).\\
The main results of our numerical study, whose technical details are discussed at the end of the paper, are reported in Fig.\ref{intro}. The BKT critical temperature, used here to rescale properly the temperatures, has been computed by means of the Nelson-Kosterlitz\cite{KN} universal relation:  
\be
J_s(T_{BKT})=2 T_{BKT}/\pi,
\label{Jump}
\ee 
which indicates the critical point at which the superfluid-stiffness jump is expected to occur.\\ 

\noindent
This study, largely discussed in\cite{me}, have revealed that for uncorrelated disorder the Harris criterion\cite{Harris} is guaranteed not only at the critical point, but even away from it.
Indeed, not only the superfluid-stiffness jump remains as sharp as in the homogeneous case (green curve in Fig.\ref{intro}(a)), but even the low-temperature trend before the jump is unchanged once that the $T=0$ suppression of the stiffness is accounted by rescaling the curve with $J_s(T=0)$. On the other hand, for the RTF case the fragmentation of the SC state at strong disorder leads to a smoothening of the BKT jump (red curve in Fig.\ref{intro}(a)), which is symmetrically smeared out with respect to the expected transition, in strong analogy with the experimental observations in thin SC films\cite{fiory_prb83,lemberger_prl00,armitage_prb07,armitage_prb11, Kamlapure_2010,mondal_bkt_prl11,goldman_prl12,yazdani_prl13}.\\ This result has been explained in terms of an unconventional vortex-pairs nucleation in the granular SC state. Indeed, the formation of large regions with low couplings $J_{ij}$ (Fig.\ref{intro}(b)), allows the system to nucleate several vortex-antivortex pairs alread well below $T_{BKT}$, leading to the superfluid-stiffness suppression.

\noindent
In the present manuscript we want to investigate more closely these results in order to understand whether the correlated disorder changes or not the universality class of the BKT transition.


\section{Results}

\begin{figure}[h!]

\includegraphics[width=0.9\textwidth]{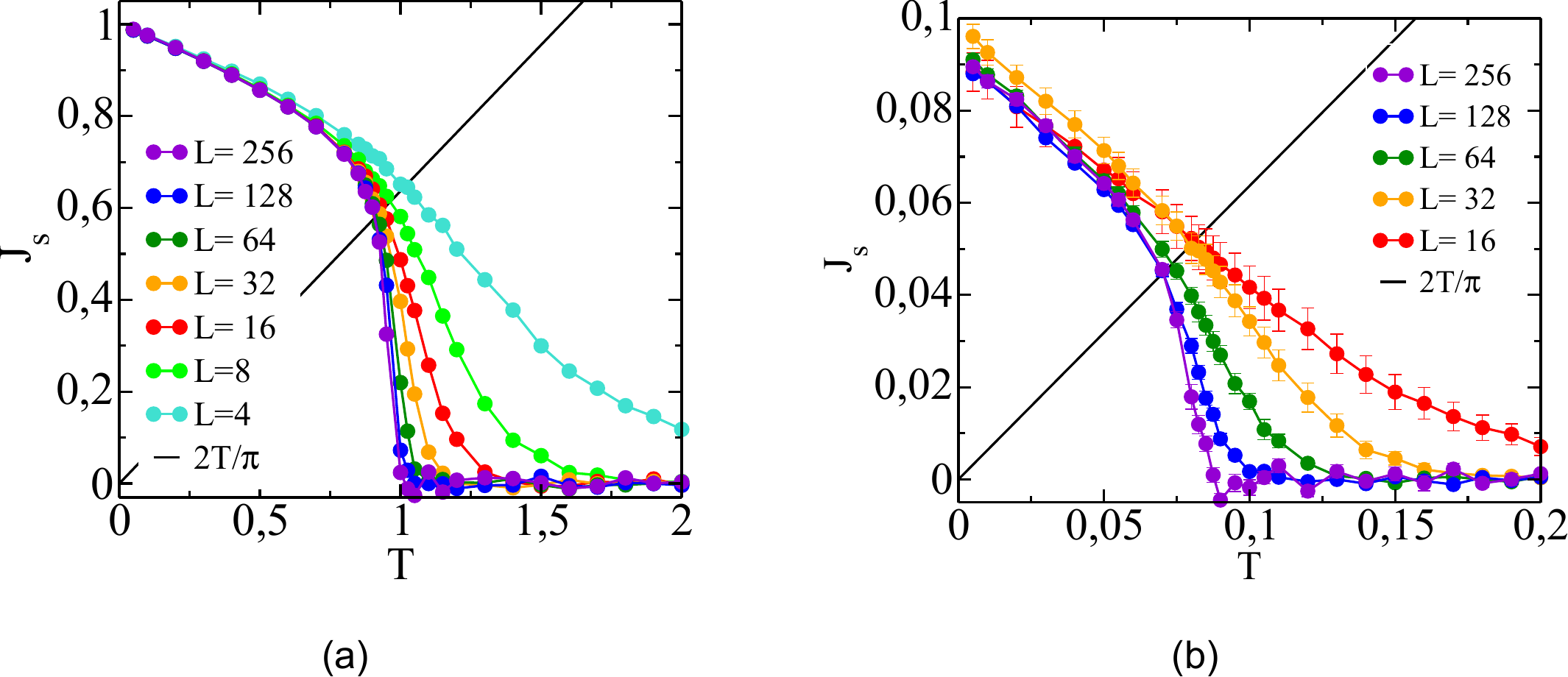}
\caption{Temperature dependence of the superfluid stiffness for different values of the linear sizes $L$. The panel (a) corresponds to the homogeneous case, while panel (b) to the disordered $RTF$ case at $W/J=10$. The solid black line in both the panels is the critical line $2T/\pi$ whose intersection with $J_s(T)$ would correspond to the critical point where the superfluid-stiffness jump is expected to occur.  }
\label{JsL}
\end{figure}

\noindent
In order to investigate the critical properties of the model \eqref{Hxy}, we need to extrapolate the thermodynamic behavior of the system via a proper finite-size scaling analysis. In the following, we will compare the well known homogeneous case ($J_{ij}=1;\,\, \forall i, j$) with the RTF model for a given disorder level, fixed here to $W/J=10$. \\
First of all, as shown in Fig.\ref{JsL}, the superfluid-stiffness jump expected in the thermodynamic limit is approached very slowly as a function of the size both for the homogeneous and for the RTF disordered case. This is exactly what one would expect from a BKT physics since the correlation length $\xi$, instead of diverging as a power-law for $T->T_c^+$, diverges exponentially as:
\be
\xi(T>T_{BKT})\sim e^{a/(T-T_{BKT})^{1/2}}.
\label{xi}
\ee
As a consequence, since finite-size effects become relevant when $\xi\sim L$, the finite-size correction to $T_{BKT}$  decrease logarithmically with the linear size of the system $L$:

\be
T^*(L)-T_{BKT} \sim \frac{1}{2\ln L} 
\label{Tshift_L}
\ee
\noindent
It appears also clear from Fig.\ref{JsL} that the superfluid stiffness scales differently with $L$ below and above $T_{BKT}$. Following\cite{Manousakis} we will study separately the two different regimes: $T<T_{BKT}$ in the first part of this section and $T>T_{BKT}$ in the second one.

\noindent


\subsection{Scaling from $T \to T_{BKT}^-$}

\begin{figure}[h!]
\centering
\includegraphics[width=0.9\textwidth]{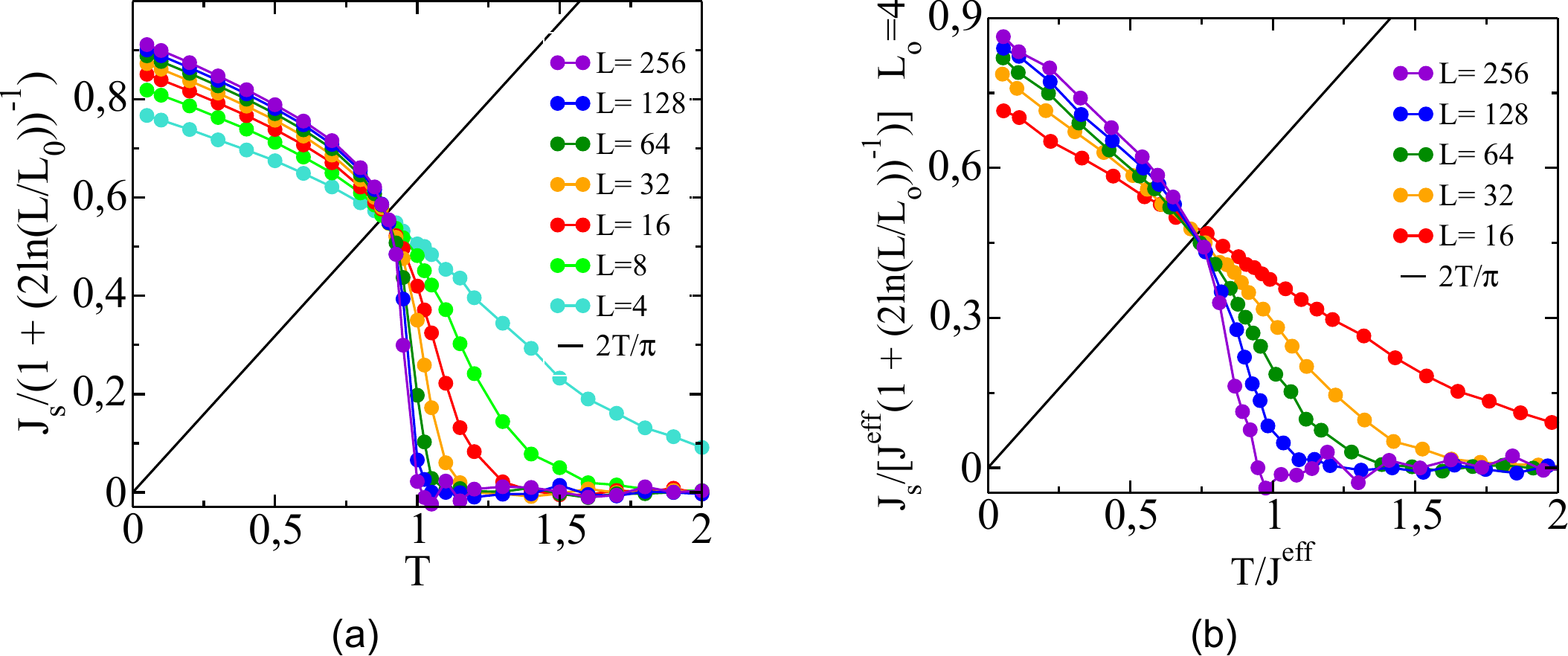}
\caption{Rescaling of the superfluid stiffness curves by means of Eq. \eqref{scalingTc} both for the clean case (a) and for the $RTF$ disordered case with $W/J=10$ (b). In the presence of disorder, for a better comparison with the clean case, one can rescale both the superfluid stiffness and the temperature by $J_{eff}= J_s(T=0)$.}
\label{Cross}
\end{figure}

\noindent
The scaling of $J_s$ in the homogeneous XY model below $T_{BKT}$ have been discussed in several papers\cite{HasenbuschTBKT, Manousakis, Sandvick}. They essentially follow from the analysis of the perturbative $RG$ equations near the BKT critical point:

\be
\begin{cases}
\frac{dx}{dl}=& -y^2\\
\frac{dy}{dl}=& -xy
\label{eqx2}
\end{cases}
\ee
where $x= \pi J_s(T)/T -2$ is the rescaled coupling constant and $y= 4\pi e^{-\beta \mu }$ the vortex fugacity, with vortex-core energy $\mu$.
When the critical line $x^2-y^2=A^2$ is approached from below $T \to T_{BKT}^-$ ($A\to 0^+$), the solution for the coupling $x$ is simply\cite{HasenbuschTBKT}:
\be
x= \frac{1}{l+c},
\label{solx}
\ee
where $c$ is a constant connected with the initial values of the $RG$ flow and $l= \ln(L)$. 
From Eq.\eqref{solx}, by the use of the Nelson-Kosterlitz universal relation\cite{KN}:
\be
J_s(T_{BKT})= 2 T_{BKT}/\pi,
\label{KN_eq}
\ee
we can derive the dependence between the finite-size value of the superfluid stiffness and its thermodynamic limit at the critical point:

\be
J_s(\infty, T_{BKT})= \frac{J_s(L, T_{BKT})}{\Big( 1+ (2\ln(L/L_0))^{-1}\Big)}
\label{scalingTc}
\ee
This means that, by rescaling the superfluid stiffness with Eq.\eqref{scalingTc}, all the rescaled curves corresponding to different $L$ will assume the same value at criticality. As a consequence, the crossing point of all them will indicate the thermodynamic value of the critical temperature itself.\\
The rescaled $J_s(T)$, obtained by tuning the value of the parameter $L_0$ in Eq.\eqref{scalingTc} in such a way to obtain the best crossing point at finite temperature, are shown in Fig.\ref{Cross}.\\

\noindent
This procedure allows us to derive the critical temperature of the RTF disordered model as well as its critical line. For the homogeneous case (Fig.\ref{Cross}(a)) the best crossing point is obtained with $L_0=1.4$, from which we extrapolate $T_{BKT}\simeq0.89$ in good agreement with\cite{HasenbuschTBKT}.
On the other hand, for the RTF disordered case ( Fig.\ref{Cross}(b)), the best crossing point has been found for $L_0=4$. The first result to be highlighted is that the presence of correlated disorder does not change the universality class of the $XY$ model, being the crossing point still on the critical line $x=0 \implies J_s(T_{BKT})= 2 T_{BKT}/\pi $. However, despite having rescaled both the superfluid stiffness and the temperature with respect to $J_s(T=0)\equiv J_{eff}$ (for the clean case: $J_{eff}=1$), the RTF disorder does change the critical temperature of the rescaled model to a lower value:
\be
T_{BKT}^{RTF}\simeq 0.71 \,J_{eff}
\label{TcRTF}
\ee
Quite interesting, this result can be physically interpreted in terms of a decrease of the effective vortex-core energy $\mu$, due to the presence of spatially correlated disorder.
Indeed for the homogeneous system it is well known\cite{benfatto_review13} that a small $\mu$ implies a larger renormalization of $J_s$ before the transition, and as a consequence a smaller value of the critical temperature.
Another remarkable effect of the presence of spatially correlated disorder is the magnification of the finite-size effects with respect to the homogeneous case. For instance, the curve of the clean case in \ref{cross_clean} relative to $L=8$ shows a tail similar for extension to the one correspondent to $L=64$ of the disordered case see \ref{cross_RTF}, hence eight times bigger than the homogeneous case.
This result is due to a larger $L_0$ scaling parameter in Eq\eqref{scalingTc} which makes the finite-size effects stronger for the disordered case.

\subsection{Scaling from $T > T_{BKT}$}

\noindent
In the high temperature regime the thermodynamic limit of the superfluid stiffness is obviously zero. The finite size effects in this region are essentially due to the correlation length $\xi$, whose divergence for $T\to T_{BKT}^+$ is cut off by the system size $L$. By means of the finite-size scaling hypothesis\cite{Fisher_Scaling, Sandvick}, we can write the rescaled superfluid stiffness as a function of the ratio between $L$ and $\xi$:

\be
\frac{J_s(L, T)}{\Big( 1+ (2\ln(L/L_0))^{-1}\Big)}= F(L/\xi)
\label{Fscaling_JsL}
\ee
\noindent
Taking thus the logarithm of the argument of the scaling function $F(x)$, Eq.\eqref{Fscaling_JsL} can be written in terms of another function $g(\ln(L/\xi))$, so that:
\be
\frac{J_s(L, T)}{\Big( 1+ (2\ln(L/L_0))^{-1}\Big)}= g(\ln(L/\xi))= g(\ln L - a/(T-T_{BKT})^{1/2})
\label{Fscaling_JsL2}
\ee

\noindent
Hence, the rescaled superfluid stiffness will have the same functional dependence on $\ln L - a/(T-T_{BKT})^{1/2}$ for each value of the system size considered.

\begin{figure}[h!]
\centering
\includegraphics[width=0.9\textwidth]{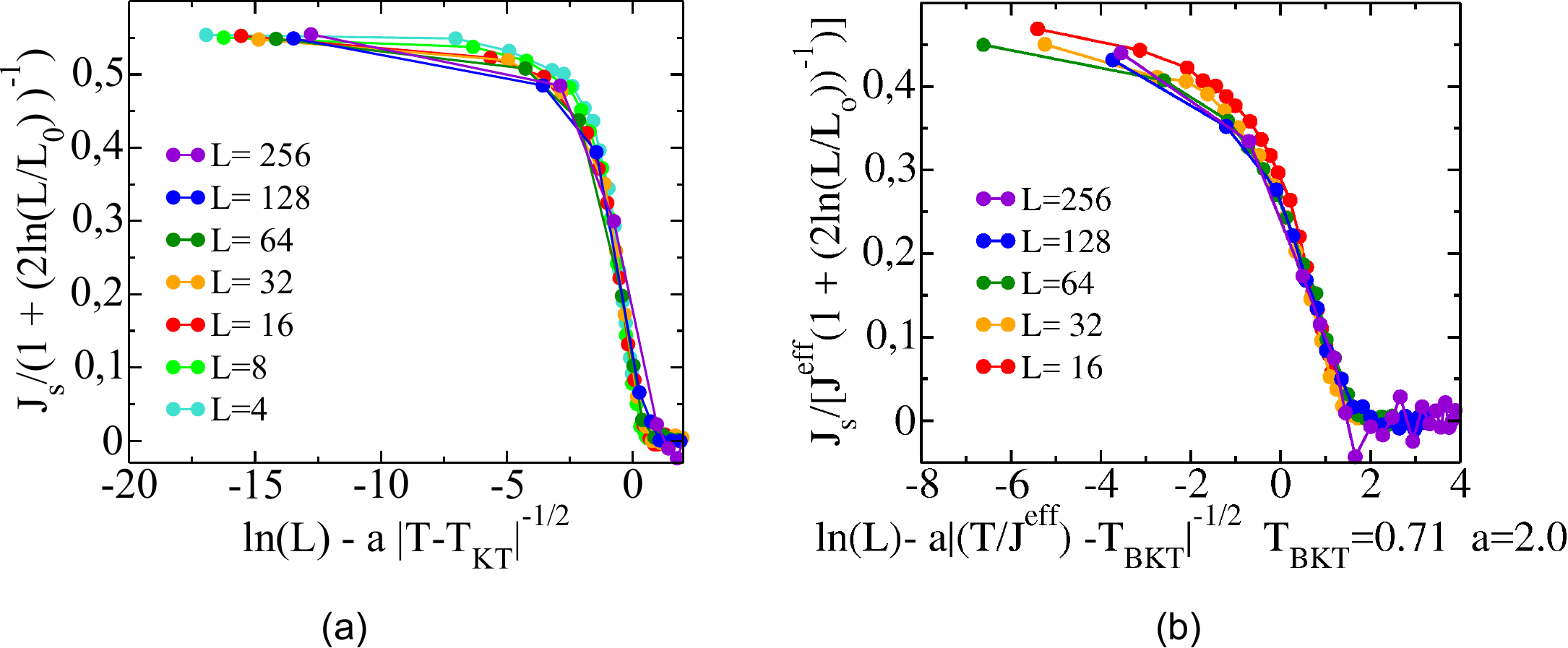}
\caption{Superfluid stiffness curves of different linear size $L$, renormalised and collapsed on the same universal curve relative to the high temperature regime: $T>T_{BKT}$. Clean case (a) and RTF disordered case with $W/J=10$ (b).}
\label{scaling}
\end{figure}

\noindent
The collapsed curves of the rescaled stiffness, obtained from our numerical data, are shown in Fig.\ref{scaling}, where we have obviously used $T_{BKT}=0.89$ for the clean case and the previously-derived  critical temperature $T_{BKT}^{RTF}=0.71J_{eff}$ for the RTF disordered case.\\
The parameter to be fixed in this study is the factor $a$ in Eq.\eqref{xi}, choosen in such way to obtain the best collapse of all the curves. For the clean case it is known\cite{Sandvick} to be $a=1.5$, while in the presence of correlated disorder we have obtained the best collapse for: $a=2.0$. The increase of the parameter $a$, by means of the presence of corraleted disorder, reflects in the scale of the x-axis, much smaller in the $RTF$ case Fig.\ref{scaling}(b) with respect to the homogeneous one Fig.\ref{scaling}(a).\\ This physically means that the correlations length $\xi$ diverges faster in the presence of correlated disorder than without it, in agreement with the previously observed increase of the finite size effects (Fig.\ref{Cross}). Let us also highlight that from the limit $T\to T_{BKT}^+$ in Fig.\ref{scaling}, which in terms of the function $g(L/\xi)$ corresponds to $\ln L - a/(T-T_{BKT})^{1/2} \to -\infty$, we can extrapolate the value of the superfluid stiffness expected  at the critical point.\\ Both for the clean and the disordered case it confirms the Nelson-Kosterlitz relation \eqref{Jump}:

\bea
J_s(\infty, T_{BKT})&=& \frac{2}{\pi}T_{BKT} \simeq 0.6\\
\frac{J_s(\infty, T_{BKT}^{RTF})}{J_{eff}}&=& \frac{2}{\pi}\frac{T_{BKT}^{RTF}}{J_{eff}} \simeq 0.45
\eea
as expected since both the critical points are crossed by the universal line $x=0$.\\
Hence, this study confirms that in both the cases the universality class is  the BKT one, showing at the same time that also in the presence of correlated disorder the correlation length diverges exponentially in the reduced critical temperature. Finally, it sheds light on the two main differences, with respect to the clean case, introduced by the spatially correlated disorder: the reduction of the critical temperature $T_{BKT}$ and the faster increase of the correlation length $\xi$ as $T\to T_{BKT}^+$.

\section{Methods}
\noindent
In our simulations each Monte Carlo step consists of five Metropolis spin flips of the whole lattice, needed to probe the correct canonical distribution of the system, followed by ten Over-relaxation sweeps of all the spins, which help the thermalization. For each temperature we perform 5000 Monte Carlo steps, and we compute a given quantity as an average after discarding the transient regime, occurring in the first 2000 steps. Finally the average over the disorder is done with 15 independent configurations for each disorder level. Where not shown, the errorbars are smaller than the point size.

\section{Discussion}
\noindent
The present paper completes the study started in\cite{me} on the 2D $XY$ model in the presence of spatially correlated couplings, which mimics the mesoscopic inhomogeneity experimentally observed in two-dimensional superconducting systems. 
From the finite-size scaling analysis, we have shown that the presence of disordered couplings with spatial correlations does not change the universality class of the BKT transition, affecting nonetheless both the critical temperature and the exponential divergence of the correlation length. More specifically, the critical temperature of the RTF model is found to be lower with respect to the homogeneous case, as a consequence of an effective smaller vortex-core energy. This result appears to be perfectly in agreement with the conclusions drawn in\cite{Mondal_2011}, where it was shown that for a correct identification of the typical BKT signatures in NbN thin films it is needed to account for $\mu$ values smaller than what expected from the standard homogeneous XY model. \\
Our work has also revealed that the presence of spatially-correlated disorder makes the finite-size
effects much stronger than in the homogeneous case, as reflected e.g. on the divergence of the correlation length as the transition is approached from above $T_{BKT}$. This result opens interesting perspectives for the understanding of the extended tails usually observed in the sheet resistance curves in 2D superconducting interfaces\cite{Bid_PRB94}. Indeed, since the resistivity goes to zero as $\xi^{-2}$ in the BKT transition\cite{HN}, the enhanced finite-size effects found in our numerical study could provide a microscopic derivation for the phenomenological models of inhomogeneity proposed so far\cite{benfatto_inho_prb09, LB_PRB84}. In addition, the presence of SC inhomogeneity at the mesoscopic level could account for the suppression of the zero-temperature stiffness with respect to its BCS estimate recently reported in $LaAlO_3$/$SrTiO_3$ \cite{LB_Bergeal}. A quantitative analysis of these effects could provide more insight into the space structure of disorder in these SC 2D materials, whose high tunability constitutes an excellent prerequisite for potential device applications.

\begin{acknowledgments}
This work has been supported by MAECI under the Italian-India collaborative project SUPERTOP-PGR04879
\end{acknowledgments}

\end{document}